\documentclass[9pt,reqno]{article}

\usepackage{geometry} 
\usepackage{graphicx}
\usepackage{enumerate}
\usepackage{amsmath,amssymb}  
\usepackage{bm}  
\usepackage{pdfsync}  
\usepackage{placeins,url}
\usepackage[T1]{fontenc}
\usepackage[utf8]{inputenc}
\usepackage{authblk}

\usepackage[compress,authoryear]{natbib}
\title{A coin vibrational motor swimming at low Reynolds number}
\author[1]{Alice C. Quillen\footnote{\texttt{alice.quillen@rochester.edu}}}
\author[2]{Hesam Askari\footnote{\texttt{askari@rochester.edu}}}
\author[2]{Douglas H. Kelley\footnote{\texttt{d.h.kelley@rochester.edu}}}
\author[1,3,4]{Tamar Friedmann\footnote{\texttt{tfriedmann@smith.edu}}}
\author[1]{Patrick W. Oakes\footnote{\texttt{poakes@rochester.edu}}} 
\affil[1]{\small Dept. of Physics and Astronomy, University of Rochester, Rochester, NY, 14627, USA}
\affil[2]{Dept. of Mechanical Engineering, University of Rochester, Rochester, NY, 14627, USA}
\affil[3]{Dept. of Mathematics, University of Rochester, Rochester, NY, 14627, USA}
\affil[4]{Dept. of Mathematics and Statistics, Smith College, Northampton, MA, 01063, USA}

\begin{document}
\maketitle

\begin{abstract}

Low-cost coin vibrational motors, used in haptic feedback, exhibit rotational internal motion inside a rigid case. Because the motor case motion exhibits rotational symmetry, when placed into a fluid such as  glycerin, the motor does not swim even though its oscillatory motions induce steady streaming in the fluid.  However, a piece of rubber foam stuck to the curved case and giving the motor neutral buoyancy also breaks the rotational symmetry allowing it to swim.  We measured a 1 cm diameter coin vibrational motor swimming in glycerin at a speed of a body length in 3 seconds or at 3 mm/s.     The swim speed puts the vibrational motor in a low Reynolds number regime similar to bacterial motility, but because of the oscillations
of the motor it is not analogous to biological organisms. Rather the swimming vibrational motor may inspire small inexpensive robotic swimmers  that are robust as they contain no external moving parts.
A time dependent Stokes equation planar sheet model  suggests that the swim speed depends on a steady streaming velocity $V_{stream} \sim Re_s^{1/2} U_0$  where $U_0$ is the velocity of surface oscillations, and streaming Reynolds number $Re_s = U_0^2/(\omega \nu)$ for motor angular frequency $\omega$ and fluid
kinematic viscosity $\nu$.


\end{abstract}



\section{Introduction}

Locomotion mechanisms of small biological organisms can inspire strategies for achieving efficient locomotion 
 in small artificial or robotic mechanisms. 
 Alternatively new construction principles can be invented that might be simpler, more efficient or more 
 more practical from an engineering perspective. 
Self-propulsion forces arise from the mechanical dissipative interactions of the organism or locomotor with a surrounding fluid, granular material or with a solid substrate (e.g., \citealt{alexander03,lauga09,lauga11,childress11,elgeti15,hosoi15,lauga16}).
Small or microscopic synthetic swimmers could be valuable in diverse fields.   For example, 
 robotic swimmers could
transport cargo, e.g., in medicine or microfluidic chips, 
or locate and process toxic materials in the environment.  

A series of motions  
of minimal complexity that can give a net body displacement often involves periodic shape changes.
At low Reynolds number in a fluid, locomotion is not possible if the body can only deform with one degree of freedom.
The Navier-Stokes equation in this limit is time independent and so trajectories (in the space of body shapes) that
are time reversible return the body to its original shape, orientation and position \citep{purcell77}.  
This implies that a scallop or clam cannot swim by open and closing its shell and that additional degrees
of freedom are required for locomotion.
This rule is known as Purcell's `scallop theorem' \citep{purcell77,lauga11}.
After a cyclical sequence of body deformations that returns the body to its original shape 
(a gait or a swimming stroke), the body translation
 and rotation depends quadratically on the amplitude of the deformations \citep{taylor51,lighthill52,shapere89,ehlers11}.
Transformations of the body shape are elements of a symmetry group that is embedded
in a larger manifold that includes body translations and rotations. 
In this sense swimming at low Reynolds number can be considered a gauge theory \citep{shapere89}.
In the language of differential geometry,
locomotion is only possible if the infinitesimal generators of deformations do not commute \citep{purcell77,shapere89}.
Constraints on the body arising from its interaction with the external world give rise to
a {\it connection} on the principal bundle of shapes and the holonomy of a closed loop in body shape space
give rise to the net motion of a swimming stroke.  

Appendages such as cilia and flagella present both manufacturing and operational challenges
for microscopic robots \citep{hogg14}. They could break, fall off, get stuck or damage tissues.
Some bacteria swim without flagella or cilia, due to traveling wave-like deformations along their cell surfaces \citep{ehlers96,ehlers11,harman13}.
 Spirochetes generate thrust by rotating long helical flagella. However, spirochete flagella are internal, residing within  the space between the inner and outer membranes and the flagella are never in direct contact with the external fluid \citep{harman13}.  Because they swim without cilia or flagella, spirochetes could inspire strategies for robotic swimming
 without appendages at low Reynolds number.

At low Reynolds number, the Navier Stokes equation becomes
 $\mu \nabla^2 {\bf u} = \nabla p$,  (the Stokes equation)
with $p$ the pressure, $\mu$ the 
viscosity and ${\bf u}$ the velocity.  
The Stokes equation combined with a condition for an incompressibility $\nabla \cdot {\bf u} =0$
is known as Stokes flow.
One can solve this elliptic partial
differential equation  (see \citealt{stone96}),
1) with any choice of boundary velocities that produce the instantaneous change of shape,
2) by subtracting an appropriate instantaneous, infinitesimal rigid motion counter flow so that the net force and torque on the body from the combined flows vanish.
%
There is a related connection between the locomotion of a body due to oscillatory motion and
the locomotion of a body due to oscillations in the fluid.
\citet{vladimirov13} proposed that a vibrating dumbbell micro-bot, comprised of
two spheres of different density separated by a vibrating actuator, could swim in a viscous fluid.
\citet{klotsa15} later showed experimentally that
two unequal density spheres connected with a spring swam when placed in a vibrating viscous fluid.
\citet{nadal14} explored propulsion of a nearly round particle that is floating in a vibrating fluid.
Swim velocities, motivated by biological organisms, have primarily been estimated
for Stokes flow (e.g., \citealt{ehlers11}), though \citet{khapalov13} studied control of a system of connected
parts using the time dependent or non-stationary Stokes flow equation; 
$\rho \frac{\partial {\bf u}}{\partial t} = - \nabla p + \mu \nabla^2 {\bf u} $, with $\rho$ the fluid density.

A simple and inexpensive vibrational swimmer would provide opportunities to explore experimentally
the hydrodynamics of locomotion for oscillating and vibrating mechanisms.  
Such a mechanism would aid or inspire development of 
small and robust robotic swimmers. 

In this paper we present a novel and low-cost swimmer constructed from a coin vibrational motor.  In section \ref{sec:swimmer}  we describe the internal rotational motion
of a coin vibrational motor and the construction of a swimmer.
Using a movie of the swimmer (see \url{https://youtu.be/0nP2MgzaOyU})  in section 
\ref{subsec:dims} we measure the swim speed  in 
glycerin and tabulate dimensionless parameters describing
the hydrodynamics regime.  In section \ref{subsec:particles} particle trajectories are used to
illustrate  and measure induced fluid motions.   Internal oscillations of the vibrational
motor cause periodic motions of the rigid motor case and this induces steady streaming in
the fluid.   The streaming motions are predominantly rotational about the motor.
To swim, the rotational symmetry of the mechanism and associated flow must be broken, 
and this is achieved with a piece of foam rubber
that also maintains neutral buoyancy and prevents the motor case from rotating. 
Within the context of the physical quantities
measured from our movie, in section \ref{sec:hydro} we discuss the hydrodynamics.  
In section \ref{subsec:planar}
we modify the planar sheet model by \citet{taylor51}. 
Using the time-dependent
Stokes equation we compute the steady streaming velocity induced
 by the oscillatory motion of a rigid planar sheet.
This and a rough application of the tangent plane approximation gives
us an estimate for the motor swim velocity and provides
an explanation for why the vibrational motor swims.
A summary and discussion follows in section \ref{sec:sum}.

\begin{figure}
\centering\includegraphics[width=2.5in]{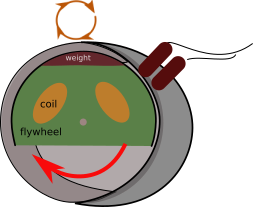} 
\caption{An illustration of a coin vibrational motor.  The motor case is rigid and we assume that its rotation
is prevented.  
A lopsided flywheel inside rotates
causing the entire case to move in a circle in a direction countering the motion of the internal weight.  
Each point on the surface of the case moves in a circle, here shown with a brown circle with arrows.  
The case surface 
 translates first downwards, then to the left, then upwards, and then to the right (or vice versa
if the flywheel inside the motor rotates in the opposite direction).}
\label{fig:cartoon_internal}
\end{figure}

\begin{figure}
\centering\includegraphics[width=3.0in]{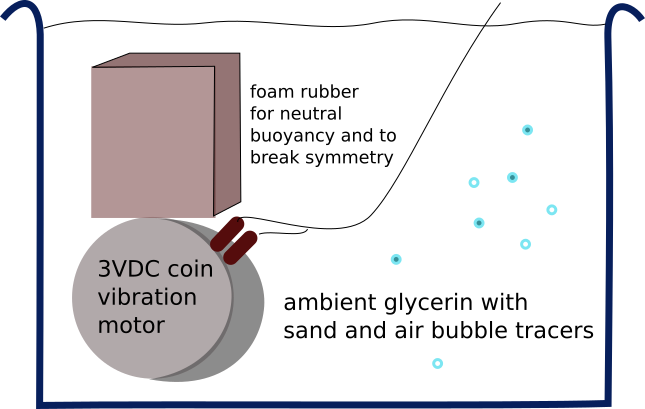} 
\begin{center}
\caption{An illustration of a swimming vibrational motor.  Foam rubber is used to make the
swimmer neutrally buoyant and break rotational symmetry.  The actual container used is much larger
than the one illustrated here, relative to the size of the motor.}
\label{fig:cartoon}
\end{center}
\end{figure}

\section{A vibrational motor swimmer \label{sec:swimmer}}

Slim and compact coin vibrational motors are cheap and ubiquitous because of their use 
in cell phones and pagers.  Electric rotating mass (ERM) vibrational motors
vibrate because they contain a lopsided internal flywheel that rotates, at typically between 10000 and 12000 rpm;
see Figure \ref{fig:cartoon_internal} for an illustration.
As the moving object is internal to the case, vibrational motors provide a novel and low-cost 
realization of the robotic swimmer concepts
proposed and investigated by 
\citet{childress11,ehlers11,vetchanin13,vetchanin16}, who  considered locomotion in  fluid
of an idealized body containing an internal mass that moves within the body (also see
\cite{saffman67}).  
Locomotion due to motions internal to the body at the cellular level was investigated by \citet{gonzalezgarcia06}.
A bacterial analog might be the spirochete where the rotating flagella lie beneath an outer
cellular membrane, though there the internal motion causes deformation of the outer membrane and the body 
is approximately incompressible. In contrast, 
the case of the vibrational motor is rigid but the mass density inside is not homogeneous.

We use a 3VDC coin vibrational motor (digikey part number 1597-1244-ND, 
manufacturer Seeed Technology, manufacturer part number 316040001, price \$1.20)
with diameter 10~mm and width 2.5~mm.   
The internal flywheel is specified  to run at least at 10000 rpm 
(according to the specification sheet provided by the manufacturer)
corresponding to a frequency of 167 Hz
and an angular frequency of $\omega = 1050 \ {\rm s}^{-1}$.  
The frequency, 167\ Hz, is barely audible but ideal
for haptic feedback.

We powered the motor  with  a table top regulated DC power supply.
The power is connected to the motor with 42 AWG polyurethane coated magnet wire (0.064 mm diameter).
Fine wire was chosen 
so as to minimize drag from the power wires as they move through the fluid.

 In free space, the geometric center of the 
case would traverse a circle centered on the system's center of mass, and rotating
in a direction opposite to that of  the lopsided flywheel. However,
the foam rubber and power wires 
 stop the rigid motor case from counter-rotation but not from moving altogether.
 Each point on the case surface moves in a small loop (see Figure \ref{fig:cartoon_internal};  assuming
 that the case is not rotating).
A rotation period consists of first moving upward, then to the right, then downward then to the left and then back
to the original position (or vice versa if the flywheel rotates in the opposite direction).  Each point on the case surface executes the same motion simultaneously.
Taking a cylindrical coordinate system with origin at the center of mass, we consider a moment
when the case is moving upward.  At that time,
 the direction of motion is  normal to the 
surface at a point on the top of the case,  normal to the surface but in the opposite direction
at a point on the bottom of the case.  However, the direction of motion is tangental to the surface
at points on the left and right sides of the case.  Tangential wavelike surface motions have been shown to be particularly
efficient at causing locomotion at low Reynolds number \citep{delgado02,gonzalezgarcia06}.
A sphere that oscillates both radially and moving back and forth along a particular axis excites larger
streaming motions than if executing one of these motions alone \citep{longuethiggins89}.

At a single moment in time the motor case moves in a single direction.   As the internal flywheel
rotates, the velocity vector of the case  rotates.   Induced surface velocities and displacements are axisymmetric
in the sense that the displacements and velocities at one time are identical to those at another time after
performing a rotation about the center of mass and shifting the phase of oscillation.
We have ignored  the location of  the power wires.
Because of the rotational symmetry of the case/fluid boundary, 
the total momentum imparted to the fluid should average to zero,
 though  the total torque exerted on the fluid might not average to zero.
If neutrally buoyant, hence without foam rubber, the object  could not swim, 
though the motor case could rotate.  

The vibrational motor is denser than water or glycerin so alone it sinks. 
Foam rubber, sponge neoprene stripping with adhesive backing, used for weather stripping,
was used both to break the rotational symmetry 
and to make it neutrally buoyant  (in glycerin).   
We attached the foam rubber  to the curved rim of the vibrational motor case
 with double sided tape near the power wires.
The rubber foam pad is about the same width as the vibrational motor so that when
viewed edge-on the entire body (motor + rubber) is about the same width ($\sim 3$ mm).
We started with a rectangular piece of rubber and then trimmed it to 
achieve neutrally buoyancy.
A cartoon of the neutrally buoyant swimmer is shown in Figure
\ref{fig:cartoon} and color photos of it are shown below (see Figure \ref{fig:twosnap}).  
The buoyancy of the foam rubber also prevents the motor case from rotating.

\begin{figure}
\centering\includegraphics[width=4.0in]{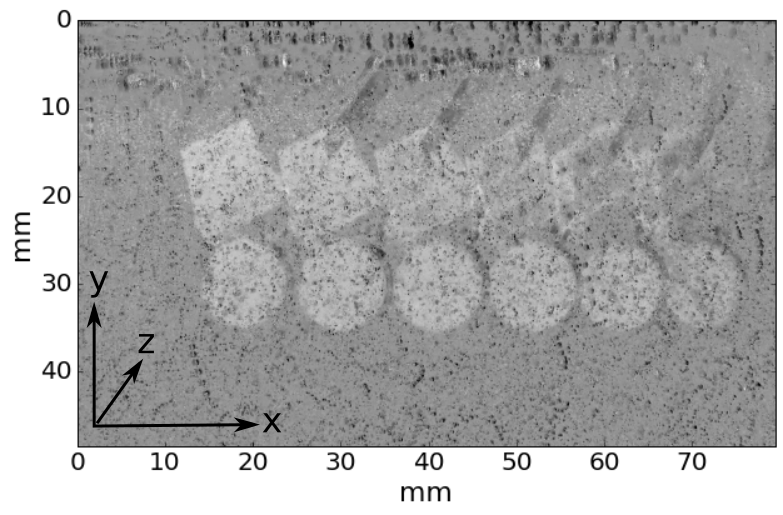} 
\begin{center}
\caption{Six movie frames, each separated by 3.2 seconds, were added after median subtraction.  This 
shows that the vibrational motor moved a body length every 3.2 seconds.  The graininess is due to sand
and air bubbles that we used to track fluid motion. The vibrational motor is moving to the right.
Coordinate axes are shown for describing the motor case oscillations (equation \ref{eqn:disp}).}
\label{fig:stack}
\end{center}
\end{figure}

The vibrational motor swimmer was placed in glycerin (99.7\% pure vegetable food grade anhydrous glycerin) in which 
we mixed some fine blue sand.  Mixing the sand also introduced some air bubbles.
The  air bubbles and fine sand particles served as tracers for the fluid motion.
We filmed the swimmer using a 1 Nikon model V1 camera with a 1 Nikkor VR 10-30 mm f/3.5--5.6 lens.
Frames from a movie are shown in Figures \ref{fig:stack}--\ref{fig:quiver} and the movie itself is available
for view at \url{https://youtu.be/0nP2MgzaOyU}.
Each movie frame has 1280$\times $720 pixels and there are 56.48 frames per second.
The movie shows that the vibrational motor swims through the glycerin at a rate of
1 body length in about 3 seconds.  
We filmed the moving vibrational motor in a container with interior $8.25 \times 13.5$ cm filled
to a height of 6.5 cm with glycerin.  We checked that the motor swam at the same speed in a container
twice as larger in all dimensions to ensure that
we don't mis-interpret boundary affects.  We also checked that the motor swam
 in clean glycerin lacking sand or air-bubbles.

When used for haptic feedback, usually the flat side of the coin vibrational motor is 
bonded to a surface that is meant to be touched or held.    
We could achieve neutral buoyancy
with the foam rubber fixed to a flat face of the motor case. 
However we found that swim speed is maximized by sticking the foam rubber to the motor's curved rim
rather than the flat face.
We infer that breaking the rotational symmetry is important for efficient locomotion in glycerin.
The simplest mechanical locomotors have a single degree of freedom for motion,
but can move if symmetry between forward and backwards strokes is broken.
For example, locomotion is achieved on a table top using a velocity dependent stroke for two masses connected
by an actuator  with asymmetric friction  \citep{wagner13,noselli14} and with 
 a single hinged microswimmer in a non-Newtonian fluid \citep{qiu14}.
For our vibrational motor, points on the case do not move along a line segment but  in a loop,   
and the rotational symmetry of the motion prevents  locomotion unless this symmetry is broken.

Rubber foam behaves visco-elastically and damps vibrations.
The rubber foam displaces fluid that could have been moving including 
near the motor surface where fluid motions are induced.
 The rotational symmetry of the motion is broken
 as both of these effects occur on only one side of the motor.

Some notes may be helpful for the reader attempting to build a similar swimmer: 
The vibrational motor cases are not designed to be impermeable to liquids.
If left unpowered in the glycerin for a few hours, when turned on later they rotate less quickly then stop working.
They are not meant for continuous operation (cell phone buzzes are short) and so can only run continuously for
an hour or two before they wear out.  
A closed-cell rubber foam, preventing absorption of fluid, is superior as the device would maintain a constant buoyancy.
Vibrational motors can be powered autonomously with a pair of 1.5V silver-oxide coin batteries 
but not with coin lithium cells.

\FloatBarrier

\begin{table}
\centering
\vbox to50mm{\vfil
\caption{ Properties of the vibrational motor swimming in glycerin  \label{tab:tab}}
\begin{tabular}{@{}lllllll}
\hline
Radius of motor & $R$ & 5 mm \\
Motor width & $w$  & 2.5 mm\\
Vibrational motor frequency &$f$  & 10000 rpm \\
Vibrational motor angular frequency & $\omega$ & 1050 s$^{-1}$ \\
Amplitude of motor case oscillations & $A$ & 0.25 mm \\
Speed of oscillatory case motions & $U_0$ & 262.5 mm/s \\
Kinematic viscosity of glycerin & $\nu$ & $1.12 \times 10^{3}\ {\rm mm~s}^{-1}$\\
Viscous diffusion length & $l_d$ & 1.46 mm\\
Swim speed & $V_{swim}$ & 3.1 mm/s \\
Swim Reynolds number & $Re_{swim}$ & 0.027 \\
Strouhal number & $St$ & 20 \\
Reynolds number & $Re$ & 1.2 \\
Streaming Reynolds number & $Re_s$ & 0.06 \\
Frequency parameter & $\beta$ & 25 \\
\hline
\end{tabular}
}
\end{table}

\subsection{Swim speed,  amplitude of motion and dimensionless quantities \label{subsec:dims}}

When powered, the vibrational motor causes motion in the fluid and the vibrational motor
swims steadily through the glycerin.  
Frames from the movie were extracted  using a command-line version of \texttt{FFmpeg}.
FFmpeg, \url{http://ffmpeg.org/}, 
 is a free software project that produces libraries and programs for handling multimedia data.
By extracting frames separated by different time intervals we measured the swim speed. 
Using the standard \texttt{scypi} Python software
package, we converted six images images, each separated by 3.2 seconds,  to grayscale and 
combined them together computing a median of the values from each frame at each pixel location.
We subtracted the median image from each image and then added the six together.
The result is shown in Figure \ref{fig:stack} and illustrates the steady swim speed 
caused by the vibrational motor.
The swim speed is 1 body length (using the motor diameter of 10 mm)
in 3.2 seconds, equivalent to  a velocity of $V_{swim} \approx 3.1$ mm/s. 

Glycerin (Glycerol) has a viscosity of 
$\mu = 1.412$ Pa~s \citep{segur51} and a density $\rho  = 1.261$ g~cm$^{-3}$ corresponding to  a kinematic
viscosity of 
\begin{equation}
\nu = \frac{\mu}{\rho} = 1.12 \times 10^{-3} \ {\rm m}^2 {\rm s}^{-1} = 1.12 \times 10^3 \ {\rm mm}^2 {\rm s}^{-1}.
\end{equation}
Taking a length scale equal to the diameter of the vibrational motor, $D=10$mm, 
and the swim velocity $V_{swim} = 3.1$ mm/s we compute the Reynolds number of the swim motion
\begin{equation}
Re_{swim} = \frac{DV_{swim}}{\nu} = 0.027.
\end{equation}
The swim Reynolds number $Re_{swim}$  can be compared to that of other swimming organisms
or mechanisms.  
The low Reynolds number implies that the vibrational motor, moving through glycerin, is in
a regime similar to bacterial motility.  However,
this Reynolds number ignores the amplitude and frequency of motor motions.

\begin{figure}
\centering\includegraphics[width=2.5in]{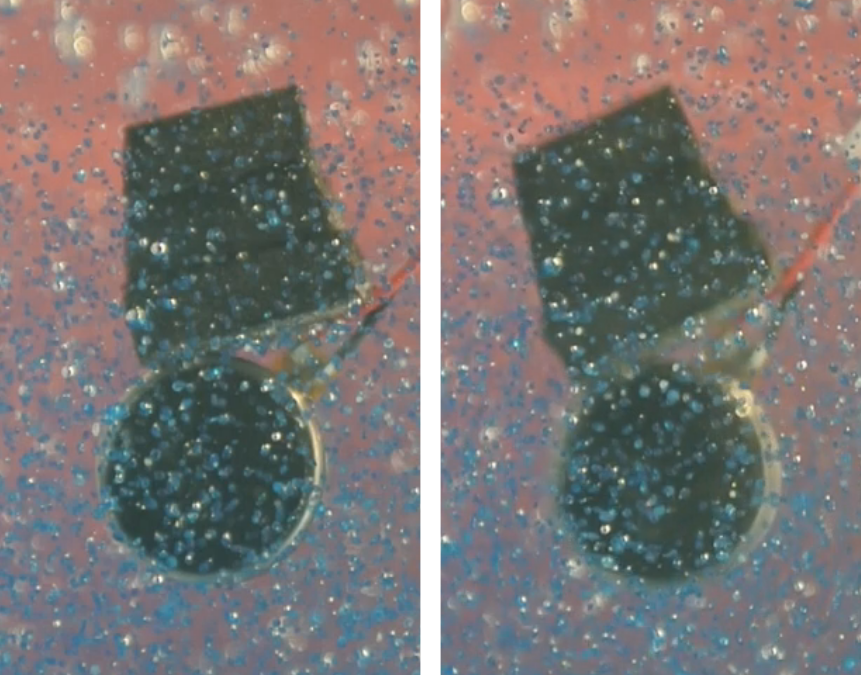} 
\begin{center}
\caption{Regions of two movie frames.  The photo on the left was taken when the motor is off.
The one on the right is when the motor is on.  The edges of the vibrational motor
on the right appear blurred because of the motor case oscillations.  We use the thickness of the bright rim
of the vibrational motor to estimate the amplitude of oscillation.
The motor is orientated at a different angle, from vertical in the right image compared
to that on the left.
The difference in orientation angle 
is from the torque on the motor caused by the rotational fluid motion.
The buoyancy of the foam rubber prevents the motor from rotating.
}
\label{fig:twosnap}
\end{center}
\end{figure}

\begin{figure}
\centering\includegraphics[width=4.0in]{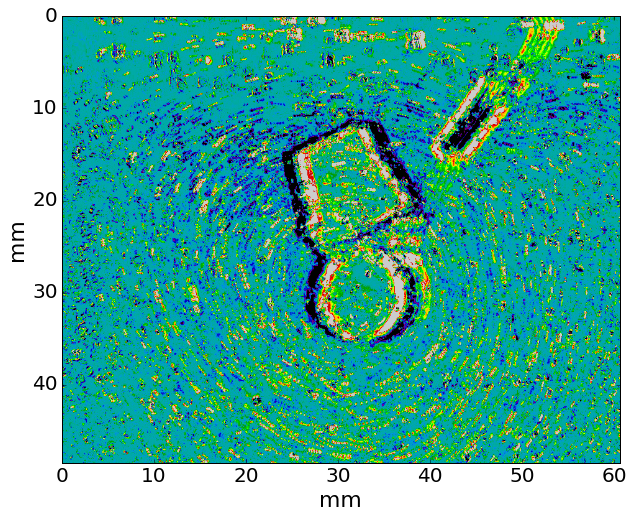} 
\begin{center}
\caption{Six movie frames each separated by 0.2 second were added after median subtraction. 
Each streak is comprised of about six particle positions for a single particle.
The streaks are not due to vibrations from the motor but rather show  motions or particles and
air bubbles embedded in the glycerin.   
 The rainbow color map was chosen to show the streaks.
The black bar on the upper right is from a piece of transparent tape we used
to insulate the electrical connections from each other.  Flow in the fluid is primarily rotational around
the vibrational motor. }
\label{fig:medstreak}
\end{center}
\end{figure}

In Figure \ref{fig:twosnap} we show regions from two movie frames. The one on the left
shows the vibrational motor at rest, before the motor is turned on.  The one on the right
shows the vibrational motor while it is on and is swimming.   The motor is blurred on 
the right due to the case oscillations.  We measured the width of the right edge of the motor
in the two frames, finding that it is 4.1 pixels wide for the image on the left
and 11.3 pixels wide for the image on the right.  
We measured the 
the pixel scale, 0.06734 mm/pixel,  using the diameter of the motor.
The increase in width of the rim is 7.2 pixels and corresponds to a distance of $0.485$ mm. 
Dividing this by two, we estimate the amplitude of surface oscillations, 
\begin{equation}
 A \approx 0.25\ {\rm mm},
 \end{equation}
 with an error of about a pixel width or $\pm 0.07\ $mm.
 Orienting a coordinate system with a motor face lying in the $x,y$ plane and $z$ along the line of
 sight (as viewed in Figures \ref{fig:cartoon_internal} - \ref{fig:quiver}, and see Figure \ref{fig:stack} to see
 the axes), 
we can describe a point on the case with mean position ${\bf x}_0 = (x_0,y_0,z_0)$ and position ${\bf x}$ as moving 
in a circle with 
 \begin{equation}
 {\bf x}(t) = {\bf x}_0 +  A \cos (\omega t) \hat {\bf x} - A \sin (\omega t) \hat {\bf y} 
\end{equation}
or with a displacement vector
\begin{equation}
{\boldsymbol \delta}  = A \cos (\omega t) \hat {\bf x} - A  \sin( \omega t)\hat {\bf y} \label{eqn:disp}
\end{equation}
that applies to each position on the motor case.
Here $\hat {\bf x} $ is a unit vector in the $x$ direction.
The displacement vector also describes the motion of the case/fluid boundary.
It is useful to describe the amplitude of motion in units of the vibrational motor's radius, $R=5$mm;
\begin{equation}
\epsilon \equiv \frac{A}{R} = 0.05.
\end{equation}
Using the amplitude we can estimate the velocity of the oscillating case surface 
\begin{equation}
U_0 = A \omega  = 262.5\ {\rm mm~s}^{-1}.
\end{equation}
The velocity of oscillation exceeds the swim velocity by a factor about 100.

Vibrating or oscillating objects in a fluid can be described in terms of dimensionless parameters;
\begin{eqnarray}
{\rm Strouhal\ number:} & St \equiv R\omega/U_0 = R/A  = \epsilon^{-1} = 20  \label{eqn:strou}\\
{\rm Reynolds\ number:}  &  Re \equiv U_0 R/\nu =  A \omega R /\nu = 1.2, \label{eqn:Re}
\end{eqnarray}
(e.g., following \citealt{kim91,riley01}). 
Combining these two parameters  we compute a parameter 
called the streaming Reynolds number \citep{brennen74,riley01},
\begin{equation}
{\rm Streaming\ Reynolds\ number:} \qquad Re_s \equiv Re \ St^{-1} = Re \ \epsilon = \frac{U_0^2}{\omega \nu} = 0.06.
\label{eqn:Re_s}
\end{equation}
The two parameters can also be combined into a frequency parameter, $\beta$,
(e.g., section 6.15; \citealt{poz11}),  sometimes called the
oscillatory Reynolds number, or a frequency Reynolds number (\citealt{childress11}),
\begin{equation}
{\rm Frequency \ parameter:} \qquad \beta \equiv Re\  St = Re/\epsilon = \frac{R^2 \omega}{\nu}  = 23.4.
\end{equation}

Important for both time dependent Stokes flow and oscillating boundary layer problems  
is the viscous diffusion length scale 
\begin{equation}
l_d \equiv \sqrt{\frac{2\nu}{\omega} }  = 1.46 \ {\rm mm}. \label{eqn:l_d}
\end{equation}
Fluid oscillations are expected to exponentially decay on this length scale  \citep{schlichting32}.
The ratio $A/l_d = \sqrt{Re_s/2}$, so the condition that the amplitude of motion is smaller than
the diffusion length, $A \ll l_d$, is equivalent to streaming Reynolds number $Re_s \ll 1$.

The properties and dimensionless parameters
of the vibrational motor swimming in glycerin are summarized in Table \ref{tab:tab}.

\subsection{Fluid motion and particle trajectories \label{subsec:particles}}

Using six frames separated by 0.2 seconds we constructed a median image and
again summed median subtracted frames.  The result is shown in Figure \ref{fig:medstreak}.
Sand particles and air bubbles embedded in the glycerin moved during the 1.2 second interval
and they appear in this figure as streaks. 
Each streak is made by a single particle and comprised of six particle positions.
For sand particle diameter 0.2 mm, 
a velocity of 8 mm/s  and kinematic viscosity of glycerin, the Stokes number is $\sim 10^{-4}$ and low
enough that the sand particles closely follow fluid streamlines. 
The black bar on the upper right is from a piece of transparent tape we used
to insulate the electrical connections from each other.  
Flow in the fluid is rotational around the vibrational motor.   The rotational streaming
lies well outside the width $l_d$ for an oscillating boundary layer, implying that 
what we are seeing is steady streaming induced by the oscillation near the motor surface.

 Figure \ref{fig:twosnap} shows that the foam rubber is tilted (to the left) when the motor
is on (photo on the right) compared to its angle when the motor is off (photo on the left).
The  rotation in the fluid  implies that there is
a torque on the motor.  Because it is more buoyant than the motor, 
the rubber foam remains on top of the case.  The buoyancy force and motor weight counteract
the torque on the motor, explaining why it is tilted when the motor is on. 
The wires also prevent rotation.

\begin{figure}
\centering\includegraphics[width=4.0in]{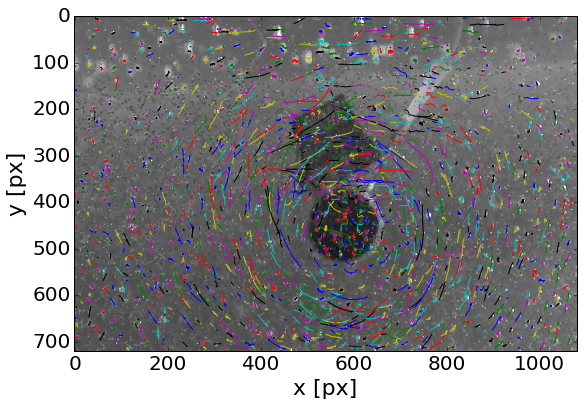} 
\centering\includegraphics[width=4.0in]{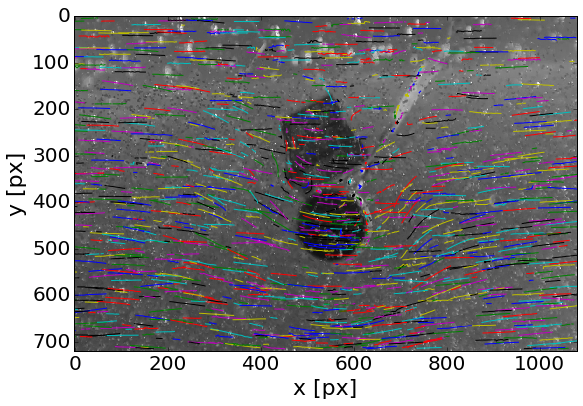} 
\caption{a) Particle trajectories in the lab reference frame are shown over a 3.5 s time interval with background image
showing a movie fame from mid-interval.  Particle trajectories are primarily concentric 
around the motor,  showing circulation.   Particle motion is clockwise.
b) Particle trajectories in the frame of the moving motor are shown over a 1.5 s time interval with background image
showing a movie fame from mid-interval.  Particles in front of the motor (the right side) are pushed down, 
under and around
the back side of the motor.
The white rectangle on the wire is a piece of transparent tape we used to insulate the wire connections from
each other.
}
\label{fig:streak1}
\end{figure}

\FloatBarrier

We used the soft-matter particle tracking software package \texttt{trackpy} \citep{trackpy} to identify
and track the air bubbles and blue sand particles that are present in the glycerin and seen in the video frames.
Trackpy is a software package for finding blob-like features in video, tracking them through time, and analyzing their trajectories. It implements and extends the widely-used Crocker-Grier algorithm \citep{crocker96} in Python.
Particles or air bubbles are identified as peaks in the grayscale image frames.
Their positions were then tracked through 70 image frames, each separated in time by 0.05 second.
Trajectories were rejected if particles moved between frames by more than 5 pixels, were lost for 
more than 3 frames and were not seen in fewer than 5 frames.
The resulting trajectories are shown on top of an image mid time interval in Figure \ref{fig:streak1}a using
3.5 seconds of video.
This figure shows particle trajectories in the lab frame.  
By shifting each frame according to the motor swim velocity
we also corrected trajectories for the swim motion of the motor.  In Figure \ref{fig:streak1}b we 
show particle trajectories in the motor frame.  In this frame particles are stationary on the front edge of the
motor and along the wire implying that we have correctly subtracted  the motor swim speed to construct
the trajectories.  The flow under and around the motor is clearer in this frame.

We argued above that the piece of foam rubber breaks the rotational symmetry, however
the flow field in Figure \ref{fig:streak1}a seems nearly  rotationally symmetric about the vibrational motor. 
When viewing the movie,  particles on the upper right are swept downwards
under the vibrational motor and left
behind it on the left.   This flow pattern is more clearly seen with particle trajectories in the motor's frame 
(see Figure \ref{fig:streak1}b).
Particle motion is reduced near the foam rubber, and is lower above the motor than below it.
The rotational streaming motion below the motor is not mirrored by a similar flow above
the motor and so the fluid has been predominantly moved from right to left, allowing the motor
to advance or swim to the right.
The foam rubber breaks the rotational symmetry.
The propulsion arises because fluid is streaming to the left below the motor.

\begin{figure}
\centering\includegraphics[width=4.0in]{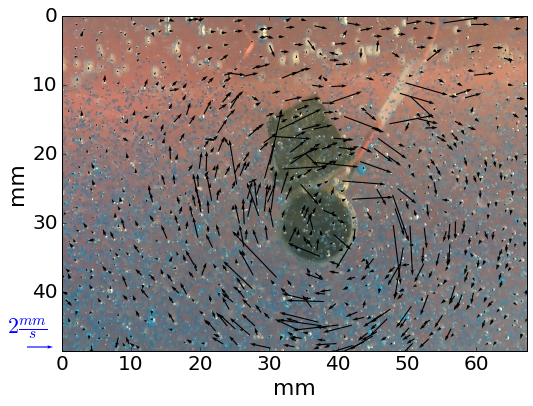} 
\centering\includegraphics[width=4.0in]{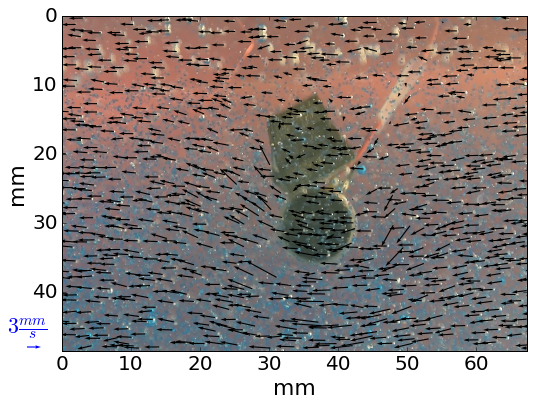} 
\begin{center}
\caption{a) Arrow lengths show particle deviations between two movie frames separated by a 0.3 time interval giving an estimate for particle velocities.  
Units for arrow length (velocity) are shown in blue on the lower left.  Here the velocities are shown in the lab frame.
b)  Particle velocities are shown in the motor's frame.
}
\label{fig:quiver}
\end{center}
\end{figure}

Particle deviations between two frames (separated by 0.3 s) were used to construct velocity vectors
and these are shown in Figure \ref{fig:quiver}a.   Velocities were estimated using the pixel
scale (0.067 mm/pixel) and the time interval between the two frames.   By subtracting the motion of the motor
we made a similar figure but for velocity vectors in the motor frame; Figure \ref{fig:quiver}b.
These figures allow us to estimate the size of the fluid velocities.
 Fluid velocities
 approximately 6 mm/s have been induced in the glycerin, with highest velocities near
 the motor surface of about 8 mm/s.
 It is somewhat clearer in the movie that the particle motions are about twice the magnitude
 of the motor swim speed.
 
 That the  flow velocities decay with distance from the motor is evident in Figure \ref{fig:quiver}a
as the arrow lengths decrease with distance from the motor. 
It is likely that the three dimensional steady velocity field could be described by the sum of 
of two Stokes flow steady singularity solutions, a rotlet and a Stokeslet  (a point force dipole)
(e.g., see section 6.6 by \citealt{poz11}, \citealt{pedley16}).  Since the flow is predominantly
rotational, the rotlet must be larger.  The rotlet speed   
decays with the square of the distance  from the vibrational motor (using Table 6.6.2 by \citealt{poz11}).

 

\FloatBarrier
\section{Hydrodynamics\label{sec:hydro}}

In the previous section we described the design and construction of a swimmer
and measurements of  quantities listed in Table \ref{tab:tab}.  
 The dimensionless parameters  the frequency parameter, $\beta$, the Reynolds number
 (computed using the oscillation velocity), $Re$, the streaming Reynolds number, $Re_s$, 
 and the Strouhal number, $St$, 
(with redundancy between the different parameters),
are used to determine which terms in the Navier-Stokes equation must be retained 
(following \citealt{kim91} chapter 6, also see section 6.15 by \citealt{poz11} or section 2 by \citealt{riley01}).   
We will use the physical sizes $R, A, U_0, \omega, l_d$ and $\nu$ 
(motor radius, oscillation amplitude, oscillation velocity, angular frequency, viscous diffusion length and kinematic viscosity) in section \ref{subsec:planar} to estimate 
 a swim speed derived from the hydrodynamics and compare it to the measured one, $V_{swim}$.  
With the measurements of Table \ref{tab:tab} in mind, 
we now discuss hydrodynamic models, first reviewing previous work.   Our
swimmer is not biologically motivated and previous hydrodynamic explorations are not quite
in the right regime.
In section \ref{subsec:planar} we modify the estimate of a steady stream velocity based on an oscillating
planar sheet by \citet{taylor51} in the Stokes flow regime.  
Using the time-dependent or non-stationary Stokes equation we compute the steady streaming velocity induced
 by the oscillatory motion of a rigid planar sheet. We then use the derived steady streaming velocity 
to estimate a swim velocity and we compare it to the one we measured from our mechanism
in section \ref{subsec:dims}. 

\citet{taylor51} considered the swimming velocity of a sheet exhibiting low amplitude traveling waves on its surface. 
Using global solutions to the Stokes equations, Taylor expanded the boundary
displacement and velocity in powers of the wave amplitude and matched expansion coefficients to solve 
for the fluid motions.
He found that the swim velocity (or induced steady streaming velocity in the fluid) 
is proportional to the square of the surface wave amplitude. 
\citet{brennen74} estimated swim velocities by computing the velocity field in an oscillating boundary layer 
in the vicinity of a surface exhibiting low amplitude plane waves. 
\citet{ehlers11} extended Taylor's formalism to more general wavelike solutions motivated by the geometrical
or gauge formalism by \citet{shapere89}.    Deformations are described in terms of a Fourier basis of vector fields.
A swim stroke is a periodic combination of these vector fields that returns the sheet (or body) to its original shape.
Curvature coefficients depend on the commutators of these vector fields and  the swimming velocity
depends on the square of the wave amplitude. 

The oscillating boundary layer of a vibrating or oscillating object in a viscous fluid should have height similar to 
$l_d$, the viscous diffusion length \citep{nyborg58,stuart66,brennen74,lighthill78}.   
However, a steady streaming motion, sometimes called
Stokes drift,  that extends past the boundary layer, can be induced by
the oscillatory flow  \citep{brennen74,longuethiggins89,riley01}.  The steady streaming flow
can be induced even in incompressible flows, so need not be called `acoustic streaming.'  
The streaming flow does not necessarily arise solely from matching the fluid velocity to a moving boundary (as by \citealt{taylor51}).
It can be caused by the Reynolds stress or advective term, ${\bf u} \cdot \nabla {\bf u}$,  
in the Navier-Stokes
equation.    Because this term is non-linear and depends on the square of the velocity, the streaming
velocity outside the Stokes layer (with width $l_d$) 
can depend on the square of the amplitude of the oscillatory motion \citep{longuethiggins89},
or $Re_s^{1/2}$ \citep{riley01},
depending on the shape and dimension of the vibrating or oscillating object and the streaming Reynolds and Strouhal numbers.

The {\it Tangent Plane Approximation} \citep{brennen74,ehlers11} 
assigns a flow velocity to each region of the surface based on expansion
of the surface deformation in terms of plane wave motions.
For a slender body such as a flagellum for which the radius
of curvature is larger compared to the diameter of the filament, a related approximation 
is known as `Resistive Force Theory' \citep{gray55}.
This approximation postulates that each infinitesimal segment is  hydrodynamically uncoupled from the others and that
the drag forces associated with normal and tangential motions are approximately proportional to the local filament velocity, with local drag coefficients those of a straight cylinder. 

The  tangent plane approximation computes local streaming velocities
 across the surface of the motor and then sums these  velocities 
 to estimate the swim velocity of a body. 
 With the goal of using such a local approximation on the vibrational motor, we consider streaming
 caused by the oscillatory motion of a rigid planar sheet.

\begin{figure}
\centering\includegraphics[width=3.0in]{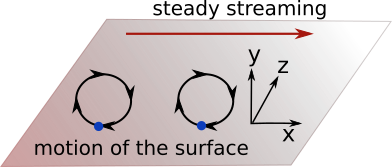} 
\caption{Illustrating the oscillation for the planar sheet.
Every point in the plane moves in a loop in $x$ and $y$ directions.  Every point moves
in the same trajectory shape simultaneously.  Here $y$ is normal to
the plane and the $x$ and $z$ directions lie in the plane.  Each point on the plane has motion
similar to every other point, so fluid motions induced by the boundary motion should be independent of $x$ and $z$.
The fluid is present in the infinite half-plane covering $y>0$ and at negative $y$ during half the oscillation.
}
\label{fig:planar}
\end{figure}

\subsection{Circular oscillation of a planar sheet \label{subsec:planar}}

We look for a solution to the equations describing the fluid that are consistent
with a rigid but moving sheet boundary.   
The motion of the vibrational motor differs from the
short wavelength (compared to body diameter) 
and traveling wave surface deformations considered by previous studies \citep{taylor51,brennen74,ehlers11}.
For our study the surface is rigid, whereas for theirs the surface deforms.
We look at a sheet with each point undergoing the same periodic motion; see Figure \ref{fig:planar} for an illustration.
The surface moves in $x$ and $y$ with the $x$ axis parallel to the sheet and the $y$ axis normal to it.
We neglect $z$ (also in the sheet plane) as there is no motion in that direction.
We use the infinite half-plane covering $y>0$.

We work in units of distance divided by the radius $R$, time in units of $\omega^{-1}$ and
velocities in units of $U_0 = A\omega$ (following \citealt{kim91} chapter 6, also see section 6.15 by \citealt{poz11}).    
We choose this scaling because we are interested
in the fluid flow at distances of $R$ rather than at distances of $A$, the amplitude of oscillation.
The surface displacement vector 
\begin{equation}
{\boldsymbol \delta} (x,t) =   \epsilon (  \cos  t \ \hat{\bf x} -  \sin t \  \hat {\bf y}) \label{eqn:boundary}
\end{equation}
where $\epsilon = A/R$
and we have omitted functional dependence on $z$ in the plane.
The velocity on the boundary is
\begin{equation}
{\bf V}_{boundary} (x,t)=  -\sin t \ \hat {\bf x} -  \cos t  \ \hat {\bf y}. \label{eqn:vboundary}
\end{equation}
Both are independent of mean $x$ value, and the mean position of the moving planar surface is at  $y=0$.
The time derivative of displacement ${\boldsymbol \delta}$ is equal to that of the velocity on the boundary
after dividing by $U_0$. 

The velocity in the fluid has components ${\bf u}(y,t) = u \hat {\bf x} + v \hat {\bf y}$.
As the motion of points in the sheet is independent of $x$, the fluid velocity  
should be independent of $x$. 
For a no-slip boundary 
the velocity of a fluid element near the surface should match that of the surface.
Using a Taylor expansion
\begin{equation}
{\bf V}_{boundary} = \left. {\bf u}\right|_{y=0} + \left. (\boldsymbol \delta (t) \cdot \nabla) {\bf u} \right|_{y=0} + ...
\label{eqn:vb}
\end{equation}
Using the y-component of the displacement (equation \ref{eqn:boundary}), 
the x-component of Equation \ref{eqn:vb} becomes
\begin{equation}
-\sin t = \left. u \right|_{y=0} - \left. \epsilon \sin t \frac{\partial u}{\partial y}\right|_{y=0} \label{eqn:stbound}
\end{equation}
to first order in $\epsilon$.
Because of our choice of units, and because $\epsilon <1$ for the vibrational motor, we can expand the
boundary condition in orders of $\epsilon$.

Equation \ref{eqn:vb} relating the fluid flow to the boundary motion
 is equivalent to equation 5 by \citet{brennen74} and  the commutator of two velocity vector fields
 shown with equation 2.19  by \citet{ehlers11}.  
The geometric paradigm is that surface motions transverse to the plane do not commute with motions
normal to the plane.

The Navier-Stokes equation 
\begin{equation}
\frac{\partial {\bf u'}}{\partial t'}  + ({\bf u'} \cdot {\boldsymbol \nabla'}) {\bf u'} =-  \frac{{\boldsymbol \nabla} p}{\rho} 
+ \nu {\boldsymbol \nabla'}^2 {\bf u'}
\end{equation}
for velocity ${\bf u}'$, time $t'$ and $\nabla' = \frac{\partial}{\partial {\bf x}'}$.
Using our rescaled variables (${\bf x} = {\bf x}'/R$, $t = \omega t'$, ${\bf u} = {\bf u'}/U_0$), the Navier-Stokes equation
\begin{equation}
\frac{\partial {\bf u}}{\partial t}  + \epsilon ({\bf u} \cdot {\boldsymbol \nabla}) {\bf u} =-  \frac{{\boldsymbol \nabla} p}{U_0^2 \rho} 
+ \beta^{-1} {\boldsymbol \nabla}^2 {\bf u}.
\end{equation}
For our problem $\epsilon \sim \beta^{-1}$ because the Reynolds number $Re \sim 1$.
In the low Reynolds number limit the smallest term is the second one on the left ($\propto {\bf u} \cdot \nabla {\bf u}$) and it can be ignored,
giving the non-stationary Stokes flow problem or linearized Navier-Stokes equation.
We do this here, but keep in mind that this is a poor approximation for our vibrational motor.
If our motor were in a somewhat higher viscosity fluid, it would be a decent approximation.
Focusing on the x-component of velocity and dropping the non-linear term
\begin{equation}
\frac{\partial u}{\partial t} =  \beta^{-1} \frac{\partial ^2 u}{\partial y^2}. \label{eqn:heat}
\end{equation}
We have neglected the pressure term because of the $x$ translational symmetry.

Equation \ref{eqn:heat} is a heat equation with  solution 
\begin{equation}
u(s) = a e^{-k_s y} \cos (k_s y - s t) + b e^{-k_s y}\sin (k_s y - s t)  \label{eqn:sol}
\end{equation}
with $k_s = \sqrt{\frac{s \beta}{2 }}$ and constant coefficients $a,b$.
We have restricted this solution so that it remains finite at large  positive $y$.
The general solution is a sum or integral over frequencies $s$.
Note that the zero frequency solution does not decay with $y$.
We consider a total solution that is a sum of a constant term ($s=0$),  a term with frequency $s=1$, and 
a term with frequency $s=2$, each with unknown coefficients.  
We insert this into our boundary condition (equation \ref{eqn:stbound})
and solve for the coefficients, finding
\begin{eqnarray}
u(y,t) \approx e^{-k_1 y} \sin (k_1 y - t)  + \frac{\epsilon k_1}{2}  
-\frac{\epsilon  k_1}{2} e^{-k_2 y}  \sqrt{2} \sin (k_2 y - 2 t + \pi/4) 
\end{eqnarray}
to first order in $\epsilon$.  Here $k_1 = \sqrt{\beta/2}$ and $k_2 = \sqrt{\beta}$.
The constant term is the streaming velocity 
\begin{equation}
u_s =  \frac{\epsilon k_1}{2} =  \epsilon \sqrt{\beta/8}  = \sqrt{Re_s/8}. \label{eqn:u_s}
\end{equation}
The steady streaming velocity is in the same direction 
as the motion of the surface at largest $y$ from the mean, (at $t=3\pi/2$ in equations
\ref{eqn:boundary}, \ref{eqn:vboundary}) and as shown with a long arrow in Figure \ref{fig:planar}.
The dependence of the streaming speed 
on $\sqrt{Re_s}$ has been seen previously in oscillating flows near a surface
(see section 2.4 by \citealt{riley01}).   In the limit of high viscosity $u_s \to 0$, the streaming velocity vanishes,  and the vibrational motor would not swim.

In plane wave studies  in the Stokes flow limit \citep{taylor51,brennen74,ehlers11}, the fluid flow also decays 
exponentially  away 
from the surface, but there the exponential decay length of the flow is set by the wavelength of surface perturbations rather
than the frequency of the motion and the viscous diffusion length at that frequency.  

Restoring units to equation \ref{eqn:u_s}
\begin{eqnarray}
V_{stream} &\approx & U_0 \sqrt{\frac{ Re_s}{8}}  = \frac{ U_0}{2} A \sqrt{\frac{\omega}{2 \nu}} \nonumber \\
&= &  \frac{U_0}{2}  \frac{A}{l_d}  =  \frac{ A \omega}{2} \frac{A}{ l_d} . \label{eqn:vstream}
\end{eqnarray}
The streaming velocity  can be written so that it is independent of radius $R$,
consistent with the planar approximation.

Evaluating the streaming speed (equation \ref{eqn:vstream}) 
for the amplitude, frequency and viscosity of the vibrational motor in glycerin (using values
listed in Table \ref{tab:tab}) we find
\begin{equation}
V_{stream} \sim 22 \ {\rm mm\ s}^{-1} . \label{eqn:vstream2}
\end{equation}
This exceeds the streaming motions we measured from particle velocities in Figures \ref{fig:quiver} 
by a factor of  3.  
We considered planar flow and the vibrational motor is only 2.5 mm wide. A better
model would take into account the width of the motor case and an associated drag force.
We approximated the flow
using unsteady or non-stationary Stokes flow, however the Reynolds number, $Re\sim 1$, is not sufficiently
low to make this approximation a  good one.  
The inaccuracy of these two approximations may account for the difference between the streaming
velocity estimated with equation \ref{eqn:vstream2} and that we measured.
The steady ($s=0$)
solution to the planar problem (equation \ref{eqn:sol}) is independent of distance from
the surface.  A three dimensional model could approximate the steady flow as
the sum of two steady singularity solutions, a rotlet and a Stokeslet 
(e.g., see section 6.6 by \citealt{poz11}, \citealt{pedley16})
with both flows decaying as a function of distance from the center of the vibrational motor.

In the second line of equation \ref{eqn:vstream} we see that the streaming velocity
depends on the 
 square of the displacement amplitude, as  expected for a second order effect.
In the low frequency limit,  the streaming velocity predicted using Stokes flow  is
$V_{swim} \sim \epsilon_\lambda^2  c$ \citep{taylor51,ehlers11},
where $c$ is the speed of surface waves and $\epsilon_\lambda = A/\lambda$
is the ratio of oscillation amplitude to wavelength.    Using the time dependent Stokes
equations (as done here)  we find a dependence on the viscous diffusion length scale
that arises because the exponential decay of the oscillating flow depends on
this scale.  In contrast in the low frequency limit, the exponential decay follows from
the solution of Laplace's equation and depends on the wavelength of the planar perturbations.
A planar model that solves the time dependent Stokes equations for plane waves should give a solution
that is consistent with both results.

Adopting the approach of the tangent plane approximation we can integrate the streaming velocities
over the motor surface, (see equation 3.1 by \citet{ehlers11} and associated discussion) 
to estimate a swim velocity. 
Velocities in front of and behind the motor
are opposite and in the vertical direction. The vertical components  don't contribute
to the horizontal swim velocity.   The velocity below the motor is unimpeded by the rubber foam
and so can be estimated from equation \ref{eqn:vstream2}. 
The motor does {\it not} get an opposite  push from its top, because the softness and 
porosity of the foam rubber 
relaxes the boundary condition (Equation \ref{eqn:vb}) and reduces the speed of the streaming motion 
on that side of the motor. 
The streaming motion on the top of the motor we can
consider to be zero.   Weighting by area and using these 4 values, 
the swim speed is about 1/4 of the streaming velocity.
We would estimate
a swim speed of a few mm/s and in the direction counter
to the flow below the motor. This is  consistent with the swim speed of the motor and suggests
that  equation \ref{eqn:vstream} is correct to order of magnitude.

Neutral buoyancy can be achieved with a light rigid body.  But if firmly attached to the vibrational motor
it too would vibrate and so induce steady streaming in the fluid.   If the rubber foam were replaced
with a solid but rigid and light body, the device would not swim because induced streaming above
the motor would counteract the induced streaming below the motor.  The foam rubber must 
absorb or damp vibrations on one side of the motor so that it will  swim.

Most estimates of swimming velocity work in the limit of small wavelength surface motions
 (wavelength smaller than the object diameter, e.g., \citealt{pedley16}) -- with swim velocities dependent on the 
 wavelength (for example equation 36 by \citealt{brennen74}).    Here we estimated
 a swim speed for a solid body motion that depends only on the velocity  of surface motion,
 frequency and fluid viscosity (equation \ref{eqn:vstream}).
The streaming motion is induced by the non-linearity of the surface boundary condition 
 (as studied by \citealt{taylor51,brennen74,ehlers11})
 rather than the advective or nonlinear term  \citep{lighthill78} because we have neglected
 the non-linear term in the Navier-Stokes equation.  

Equation \ref{eqn:vstream} suggests that the swim velocity can be maximized by increasing
the amplitude of oscillation, $A$, the angular frequency of internal rotation $\omega$ and decreasing the viscosity.
This pushes the solution to higher Reynolds number, $Re >1$, where  the advective or non-linear
term in the Navier-Stokes equation cannot be neglected.   In this regime steady streaming is still
present but the flow can be more
 complex \citep{honji81,kotas07,childress11} and perhaps 
 experimental observations or simulations are required to 
understand the flow field and resulting swim velocity.

We have assumed that the vibrational motion of the motor (equations \ref{eqn:disp}, \ref{eqn:boundary}) is circular.  
However the amplitude of case oscillation 
depends on the motor recoil and how the fluid and rubber foam opposite it.    
The surface displacement is more generally an ellipse.  

For the planar model with $\hat {\bf x}$ tangent to the surface and $\hat {\bf y}$ normal to it,
we can modify equation \ref{eqn:boundary} to describe elliptical motion;
${\boldsymbol \delta}  =   \epsilon   \cos  t \ \hat{\bf x} -  \epsilon \gamma \sin t \  \hat {\bf y}. $
The ratio of $y$ to $x$ amplitudes is  $\gamma$ and $\gamma \sim 1$ for nearly
circular motion.  Here the amplitude of motion normal to the surface is proportional to $\gamma$
and the variables are scaled to the displacement and velocity in the direction parallel to the surface.
The boundary condition (equation \ref{eqn:stbound}) and solution (equation \ref{eqn:sol}) 
to the non-stationary Stokes equation give  
 streaming velocity the same as before (equation \ref{eqn:vstream}) but multiplied by $\gamma$.
The streaming speed (including units),
\begin{equation}
V_{stream} \approx \frac{A \omega}{2} \frac{A \gamma}{ l_d}, \label{eqn:vsAAgamma}
\end{equation}
 shows the dependence on the tangential and normal amplitudes of
motion,
consistent with the geometric interpretation that the two motions do not commute.
The tangent plane approximation can again be used to estimate the streaming motions on different
sides of the motor but taking
into account how the ratio of amplitudes, $\gamma$, depends on the local angle of the motor surface.
The blur in our movie frames when the motor is on is consistent with nearly circular motion, $\gamma \sim 1$, 
so we do not modify our rough estimate for the swim velocity.

\section{Summary and Discussion\label{sec:sum}}

We succeeded in getting a 1~cm diameter coin vibration motor to swim in glycerin at 3~mm/s by breaking its 
rotational symmetry of motion.  Symmetry was broken using a piece of foam rubber that also served
to give the motor neutral buoyancy and prevent it from rotating.   
Movies of the motor swimming in glycerin with sand and bubble tracers illustrate
that rotational  steady streaming motions are induced by the motor and that the asymmetry in
distribution of
these motions is consistent with propulsion.
The swimming vibrational motor represents
a low-cost realization of the concepts for swimming by internal motions 
previously proposed and discussed  by  \citet{childress11,ehlers11,vetchanin13}.
Using the swim velocity to compute the Reynolds number, we find that it 
is low, $Re_{swim} \sim 0.03$, and in the regime of bacterial motility.
As bacteria don't vibrate, this concept for locomotion is not helpful for understanding 
bacterial locomotion.  However, our swimmer is novel and inexpensive, so
 may inspire development of small  robotic swimmers that are robust because
they have no external moving parts.

Most previous works
either focused on acoustically driven streaming due to oscillatory flows \citep{nyborg58,stuart66,riley01}
or short wavelength surface waves in the time independent (or stationary) Stokes regime to estimate
swim speeds of periodically deforming bodies \citep{taylor51,brennen74,shapere89,ehlers11,pedley16}.
We used a translationally invariant oscillating
planar motion model and non-stationary Stokes flow to estimate a steady streaming velocity.
We find that steady streaming, extending past an oscillating boundary layer, arises from a second
order effect
induced by the moving boundary, rather than by Reynolds stress (or advective non-linear term)
as is possible in many acoustic streaming problems.  Because the oscillating flow
decays exponentially with the viscous diffusion length scale, 
the steady streaming velocity is also sensitive to this scale.
Using the tangent plane approximation,
this planar problem suggests that the swim velocity should be a factor a few smaller than 
\begin{equation}
V_{swim} \sim  {A \omega} \sqrt{Re_s/8}
\end{equation}
where a correction factor  (multiplying the above estimate) 
depends on a sum of the steady streaming motions induced on different sides of the body, 
This relation is appropriate for Reynolds number $Re=U_0/(R\nu) < 1$ where $U_0=A\omega$
is the velocity of motor case oscillation.

The amplitude of motion, $A$, vibrational velocity, $U_0$, and ellipticity of motion,
depend on the force exerted on the motor by the fluid as well as the
distance between motor centroid and center of mass,  $d_m$.  
Defining a parameter $\alpha = d_m/R$ (equation 2.4 by \citealt{childress11}),  
\cite{childress11} used numerical simulations to estimate the amplitude
of motion for a linear rather than rotational shifting internal mass distribution.  Their result
is given in their equation 2.18 and is approximately consistent with  
$A/R \sim  \frac{\alpha}{2 + \beta^{-1}}$   giving $A \sim d_m/2$
for our frequency parameter $\beta \sim 25$.  The amplitude $A$ would   
need to be measured or computed  to estimate swim velocities
for different frequency vibrational motors in different fluids.  
For $\beta > 1$ and $Re\lesssim 1$, 
equation \ref{eqn:vstream} and the
scaling by \cite{childress11} suggest that the swim velocity 
$V_{swim} \sim \alpha^2 (R/l_d) R \omega/8 $.

Use of the non-stationary Stokes equation allowed us to estimate the streaming velocity
caused by a rigid but oscillating surface and is not limited to short wavelength surface perturbations
(as was true for the Stokes regime studies by \citealt{taylor51,brennen74,ehlers11,pedley16}).
However the non-stationary Stokes equation neglects 
 the advective term in the Navier-Stokes equation and this term should not  be neglected at $Re \gtrsim 1$.
 Steady streaming is predicted at this \citep{sadiq11} and higher Reynolds numbers where
the dynamics may become increasingly complex (e.g., \citealt{stuart66,honji81,kotas07,childress11}).
Unfortunately, our swimming vibrational motor, with $Re \sim 1$, is in this regime.
Also, our vibrational 
 motor is thin, with width to diameter ratio of 1/4, and the tangent plane approximation is inaccurate.
Improved calculations or simulations
can improve upon our rough approximations by taking into account the structure of the boundary
layer (or layers; \citealt{stuart66,kotas07}) and the three-dimensional structure  
of the vibrational motor and associated fluid flow (e.g., \citealt{childress11,vetchanin13}).


\vskip 2cm
Acknowledgments:
We thank Eric Blackman, Eva Bodman, Dan Tamayo, Luke Okerlund, Logan Meredith and 
Andrea Kueter-Young  for helpful discussions. 

\vskip 2cm
This is a preprint of a work accepted for publication in Regular and Chaotic Dynamics, 2016;
see \url{http://pleiades.online/}.

{}

\end{document}